\documentclass{PoS}

\title{A Numerical Unitarity Formalism for  One-Loop Amplitudes}
\ShortTitle{Numerical Unitarity Formalism}

\author{R.~Keith Ellis\\
Fermilab, Batavia, IL 60510, USA \\
E-mail: \email{ellis@fnal.gov}}
\author{Walter T.~Giele\\
 Fermilab, Batavia, IL 60510, USA \\
E-mail: \email{giele@fnal.gov}}
\author{\speaker{Zoltan Kunszt},\\
Institute for Theoretical Physics, ETH, CH-8093 Z\"urich, Switzerland\\
E-mail: \email{kunszt@itp.phys.ethz.ch}}

\def\PL #1 #2 #3 {{\it Phys. Lett.} {\bf#1} (#3) #2}
\def\NP #1 #2 #3 {{\it Nucl. Phys.} {\bf#1} (#3) #2}
\def\ZP #1 #2 #3 {{\it Z. Phys.} {\bf#1} (#3) #2}
\def\PRL #1 #2 #3 {{\it Phys. Rev. Lett.} {\bf #1} (#3) #2}
\def\PR #1 #2 #3 {{\it Phys. Rev.} {\bf#1} (#3) #2}
\def\MPL #1 #2 #3 {{\it Mod. Phys. Lett.} {\bf#1} (#3) #2}
\def\RMP #1 #2 #3 {{\it Rev.~Mod. Phys.} {\bf#1} (#3) #2}

\newcommand{\beqn}{\begin{eqnarray}}
\newcommand{\eeqn}{\end{eqnarray}}
\newcommand{\beqns}{\begin{eqnarray*}}
\newcommand{\eeqns}{\end{eqnarray*}}

\newcommand{\beq}{\begin{equation}}
\newcommand{\eeq}{\end{equation}}
\newcommand{\beqa}{\begin{eqnarray}}
\newcommand{\eeqa}{\end{eqnarray}}

\newcommand{\nn}{\nonumber \\}

\abstract{
The unitarity method for calculating one-loop amplitudes  
provides   algorithms  of polynomial complexity.
This is primarily beneficial  for the computation 
of multi-leg one loop amplitudes and  it  is therefore of great interest to develop
a numerical implementation of the unitarity method. 
We describe a recently-developed, efficient, 
semi-numerical  unitarity method for
the computation of the cut-constructible part of one-loop amplitudes.
}

\FullConference{8th International Symposium on Radiative Corrections (RADCOR)\\
		 October 1-5 2007\\
		 Florence, Italy}

\begin{document}
\section{Introduction}
At the LHC  the observation of   hard processes with many jets,
gauge bosons and perhaps heavy new particles and their subsequent
theoretical interpretation is of primary importance.   The quantitative
analysis of these processes requires the knowledge of the corresponding
cross-sections and correlations at next-to-leading order (NLO) accuracy
in perturbative QCD. The standard method of calculation
based on Feynman diagrams becomes
very cumbersome for multi-leg processes.
In gauge theories the
conventional Feynman diagram method produces intermediate
results which are  much more complicated then the final answer.
The number of Feynman graphs grows very fast with the number
of external legs. For a tree-level $N$-gluon scattering
the number of individual Feynman graphs is approximately
$N^{(N-3)}$ (within 5\% accuracy up to 16 gluons) \cite{Kleiss:1988ne}.
This decomposition generates a large number of terms.
With a growing number of external particles 
it becomes a forbidding task to simplify the resultant expression
analytically. Recently  more numerical techniques have been developed
(see e.g. ref.~\cite{Ellis:2006ss}). However, because of the 
stronger than exponential growth in the number of Feynman diagrams these
brute-force methods become
computationally very intensive for amplitudes with six or more legs. 

Unitarity methods have been suggested as an alternative, more efficient 
procedure for loop calculations  a long time ago 
\cite{Cutkosky:1960,Diagrammar,vanNeerven:1985xr}.
 Their use in
the context of gauge theories is especially beneficial~\cite{Bern:1994zx,BDDKfusing}.
The computing time is governed by
the efficiency in computing tree amplitudes and by the 
number of cuts. 
In these applications, the  unitarity cut is four dimensional.
This allows the  use of helicity method and surprisingly simple
analytic answers~\cite{Zqqgg} have been derived.  The four-dimensional unitarity
method  reconstructs only the so-called cut-constructible part of the amplitude.
The remaining rational part is obtained using known
properties of the collinear limit . 
In supersymmetric theories, which have improved ultra-violet behaviour,
the rational part vanishes.
   
Recently, new ideas   (for a  review see
\cite{Bern:2007dw})
 on  twistors~\cite{WittenTwistor}, 
multi-pole cuts (generalized unitarity)~\cite{BCFGeneralized},
recursion relations~\cite{BGRecursion,BCFRecursion,BCFW,BDKrecursionOneLoop}, 
unitarity in $D$-dimension~\cite{BernMorgan,ABFKM}
and the use of algebraic parametric integration technique~\cite{OPP}  have   made 
the unitarity cut method even more promising.
It appears that ultimately one can find an efficient algorithm
which can be used to calculate the full  one-loop amplitudes in 
terms of tree-level amplitudes.

Here we briefly  describe a semi-numerical four dimensional  unitarity method~\cite{EGK}
 which expands the algebraic method of ref.~\cite{OPP} by developing a
numerical scheme. The numerical algorithm     
evaluates only the cut-constructible part of the one-loop amplitudes.
\footnote{This   scheme  has been extended  recently into  
a semi-numerical  $D$-dimensional unitarity  method~\cite{GKM}.
For other promising methods for calculating the rational parts
see Refs.~\cite{BBDFK,BF07}.}

\section{Structure of the one loop amplitude}

\begin{figure}[tbp]
\centering
\includegraphics[width=0.48\textwidth]{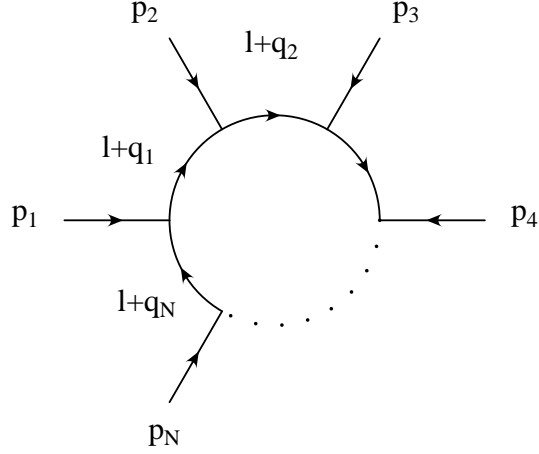}
\caption{The generic $N$-point loop amplitude. }
\label{fig:generic}
\end{figure}
The generic $D$-dimensional $N$-particle one-loop amplitude (fig.~\ref{fig:generic})
is given by~\footnote{We restrict our discussion to  (color) ordered
external legs. The extension to more general cases is straightforward. } 
\beq
{\cal A}_N(p_1,p_2,\ldots,p_N)=\int [d\,l]\ 
\frac{{\cal N}(p_1,p_2,\ldots,p_N;l)}{d_1d_2\cdots d_N}\ ,
\eeq
where $p_i$ represent the momenta flowing into the amplitude, and $[d \,l]=d^D l$. 
The numerator structure
${\cal N}(p_1,p_2,\ldots,p_N;l)$ is generated by the particle content and is a
function of the inflow momenta and the loop momentum. Since the whole amplitude
has been put on a common denominator, the numerator can also include 
some propagator factors.
The dependence of the amplitude on other quantum numbers has been suppressed. 
The denominator is a product of inverse propagators
\beq\label{parmchoice}
d_i=d_i(l)=(l+q_i)^2-m_i^2=\left(l-q_0+\sum_{j=1}^i p_i\right)^2-m_i^2\ ,
\eeq
where the 4-vector $q_0$ parameterizes the arbitrariness in the choice of loop momentum. 
The one-loop amplitude in $D=4-2\epsilon$ can be decomposed in a basis set of scalar master integrals giving
\beqa\label{MasterDecomp}
&&{\cal A}_N(p_1,p_2,\ldots,p_N)=
\sum_{1\leq i_1\leq N} a_{i_1}(p_1,p_2,\ldots,p_N) I_{i_1}\ ,
+\sum_{1\leq i_1<i_2\leq N} b_{i_1i_2}(p_1,p_2,\ldots,p_N)
 I_{i_1i_2} \nonumber \\
&+&\sum_{1\leq i_1<i_2<i_3\leq N} c_{i_1i_2i_3}(p_1,p_2,\ldots,p_N)
 I_{i_1i_2i_3} 
+\sum_{1\leq i_1<i_2<i_3<i_4\leq N} d_{i_1i_2i_3i_4}(p_1,p_2,\ldots,p_N)
 I_{i_1i_2i_3i_4}\ ,
\eeqa
where the master integrals
 are given by $\quad
 I_{i_1\cdots i_M}=\int [d\,l]\ \frac{1}{d_{i_1}\cdots d_{i_M}}\ . \quad
$
Analytic expressions for $D$-dimensional 
master integrals with massless internal lines are reported in ref.~\cite{BDDKfusing}.
The corresponding results for divergent integrals with some 
massive internal lines are reported in ref.~\cite{rep}.
The maximum number of master integrals is determined by the
dimensionality, $D$, of space-time; for the physical case this gives
up to 4-point master integrals.  The unitarity cut method is based on
the study of the analytic structure of the one-loop amplitude. The
coefficients are rational functions of the kinematic variables and
will, in general, depend on the dimensional regulator variable 
$\epsilon=(4-D)/2$.  When all the coefficients of the master integrals are
calculated in 4 dimensions we obtain the ``cut-constructible'' part of
the amplitude.  The remaining ``rational part'' is generated by the
omitted ${\cal O}(\epsilon)$ part of the master integral
coefficients.
For a numerical procedure we need to recast 
the study of the analytic properties of the unitarity
cut amplitudes into an algebraic algorithm which can be implemented numerically.
In ref.~\cite{OPP} it was proposed that one focus on the integrand of the
one-loop amplitude, 
\beq
{\cal A}_N(p_1,p_2,\ldots,p_N|l)=
\frac{{\cal N}(p_1,p_2,\ldots,p_N;l)}{d_1d_2\cdots d_N} \; .
\eeq
This is a rational function of the loop momentum.  
 We can re-express the
rational function in an expansion over 4-, 3-, 2- and 1-propagator
pole terms. The residues of these pole terms contain the master
integral coefficients as well as structures (so-called spurious terms) which reside in the ``trivial'' space,  the subspace
orthogonal to the ``physical'' space.  The  physical space is the 
subspace spanned by the external momenta of the corresponding master integral.
The spurious terms are important as subtraction terms in the determination of 
lower multiplicity poles. 
After integration over the loop momenta,
Eq.~(\ref{MasterDecomp}) is recovered. This approach transforms the
analytic unitarity method into the algebraic problem of partial
fractioning a multi-pole rational function
and allows for   a numerical implementation.

\section{Parameterization of the loop momentum on the unitarity cuts}

We restrict ourselves to a 4-dimensional space.
Given the master integral decomposition of Eq.~(\ref{MasterDecomp}) we can partial fraction the integrand
of any 4-dimensional $N$-particle amplitude as
\beq
{\cal A}_N(l)
=\!\!\sum_{1\leq i_1\leq N} \frac{\overline{a}_{i_1}(l)}{d_{i_1}}\,.
+\!\!\!\sum_{1\leq i_1<i_2\leq N} \frac{\overline{b}_{i_1i_2}(l)}{d_{i_1}d_{i_2}}\, 
+\!\!\!\!\sum_{1\leq i_1<i_2<i_3\leq N} \frac{\overline{c}_{i_1i_2i_3}(l)}{d_{i_1}d_{i_2}d_{i_3}}
+\!\!\!\!\!\sum_{1\leq i_1<i_2<i_3<i_4\leq N} 
\frac{\overline{d}_{i_1i_2i_3i_4}(l)}{d_{i_1}d_{i_2}d_{i_3}d_{i_4}}\ .
\eeq
To calculate the numerator factors, one evaluates  the residues by taking 
the inverse propagators equal to zero.
The residue has to be taken
by constructing the loop momentum $l_{ij\cdots k}$ such that 
$d_i(l_{ij\cdots k})=d_j(l_{ij\cdots k})=\cdots=d_k(l_{ij\cdots k})=0$. 
Then the residue of a function $F(l)$ is given by
\beq
\mbox{Res}_{ij\cdots k} \left[F(l)\right] \equiv
\left.\Big(d_i(l)d_j(l)\cdots d_k(l) F\left(l\right)\Big)\right\rfloor_{l=l_{ij\cdots k}}\ .
\eeq
The quadruple and triple pole residues are now given by  as 
\beq
\overline{d}_{ijkl}(l)=\mbox{Res}_{ijkl}\Big({\cal A}_N(l)\Big)\,,\quad
\overline{c}_{ijk}(l)=\mbox{Res}_{ijk}
\left({\cal A}_N(l)-\sum_{l\neq i,j,k}\frac{\overline{d}_{ijkl}(l)}{d_id_jd_kd_l}\right)\ ,
\eeq
with similar expressions for the double $\overline{b}_{i_1i_2}(l)$ and single
$\overline{a}_{i_1}(l)$ pole residues. 
As an illustration we  briefly outline how to  construct the residue functions for
quadruple cuts. 

\section{The quadrupole residue}

To calculate the box coefficients we choose the loop momentum $l_{ijkl}$ 
such that four inverse propagators are equal to zero, 
\beq\label{Resdef4}
\overline{d}_{ijkl}(l_{ijkl})=\mbox{Res}_{ijkl}\Big({\cal A}_N(l)\Big)
\quad d_i(l_{ijkl})=d_j(l_{ijkl})=d_k(l_{ijkl})=d_l(l_{ijkl})=0\ .
\eeq
We will drop the subscripts on the loop momentum in the following.
Because we have to solve the unitarity constraints explicitly, we
have to choose a specific parameterization.
In ref.~\cite{EGK}  the van Neerven-Vermaseren basis~\cite{NV} is used
which gives a very natural parameterization of the  
``trivial'' and the ``physical'' space in terms of dual vectors constructed from the  inflow momenta for
a given cut type. 
We can  decompose the loop momentum as
\beq\label{Ldef4}
l^\mu=V_4^{\mu}+\alpha_1\, n_1^{\mu}\ .
\eeq
The variable $\alpha_1$ will be determined such that the unitarity conditions $d_i=d_j=d_k=d_l=0$
  are fulfilled. $V_4$ is a well defined vector in the ``physical'' space constructed from the
three independent inflow momenta;  $n_1$ is the unit  vector of the one-dimensional trivial space.
One finds two complex solutions
\beq \label{boxtwocomplexsolns}
l_{\pm}^{\mu}=V_4^{\mu}\pm i\,\sqrt{V_4^2-m_l^2}\times n_1^{\mu}\ ,
\eeq
which are easily numerically implemented.
We note that  the four propagators are on-shell and the amplitude
factorizes for a given intermediate state into 
4 tree-level amplitudes ${\cal M}^{(0)}$. The residue of the amplitude in Eq.~(\ref{Resdef4}) 
is given in terms of tree amplitudes as
\beqa
\mbox{Res}_{ijkl}\Big({\cal A}_N(l^\pm)\Big)&=&
{\cal M}^{(0)}(l_i^\pm;p_{i+1},\ldots,p_{j};-l_j^\pm)\times
{\cal M}^{(0)}(l_j^\pm;p_{j+1},\ldots,p_{k};-l_k^\pm)\nn &\times&
{\cal M}^{(0)}(l_k^\pm;p_{k+1},\ldots,p_{l};-l_l^\pm)\times
{\cal M}^{(0)}(l_l^\pm;p_{l+1},\ldots,p_{i};-l_i^\pm)\ ,
\eeqa
where the loop momenta $l_n^{\mu}$ are complex on-shell momenta and 
there is an implicit sum over all states of the cut lines (such as e.g. particle type, color, helicity).
The tree-level 3-gluon amplitudes,
${\cal M}_3^{(0)}$, are non-zero because the two cut gluons have complex momenta \cite{BCFGeneralized}.
Any remaining dependence of the residue $\overline{d}_{ijkl}$ on the 
loop momentum enters through its component in the trivial space,
$
\overline{d}_{ijkl}(l) \equiv \overline{d}_{ijkl}(n_1\cdot l)\ .
$
The number of powers of the loop momentum $l$ in the numerator structure 
is called the rank of the integral. 
After integration we find 
that $(n_1\cdot l)^2\sim n_1^2=1$. Thus rank one is the maximum 
rank of a spurious term (which by definition vanishes upon integration over $l$).
Hence the most general form of the residue is 
$
\overline{d}_{ijkl}(l)=d_{ijkl}+\tilde{d}_{ijkl}\,l\cdot n_1\ .
$
Using the two solutions of the unitarity constraint, 
Eq.~(\ref{boxtwocomplexsolns}),
we now can determine the two coefficients of the residue
\beq\label{Cdef4}
d_{ijkl}=\frac{\mbox{Res}_{ijkl}\Big({\cal A}_N(l^+)\Big) 
+\mbox{Res}_{ijkl}\Big({\cal A}_N(l^-)\Big)}{2},\quad 
\tilde{d}_{ijkl}=\frac{\mbox{Res}_{ijkl}\Big({\cal A}_N(l^+)\Big) - \mbox{Res}_{ijkl}\Big({\cal A}_N(l^-)\Big)}{2i\sqrt{V_4^2-m_l^2}}\ .
\eeq
After the subtracting the quadruple cut contributions from the amplitude
we can repeat the procedure for the triple, double and single cuts.
 
\section{Numerical results}
As an application  in ref.~\cite{EGK} the 4-, 5- and 6-gluon scattering
amplitudes at one-loop have been  recalculated with the new method.
The cut-constructible parts of the ordered amplitudes are also known analytically 
making a direct comparison possible. 
 Also, the 6-gluon amplitude was numerically evaluated
using the integration-by-parts method~\cite{Ellis:2006ss}.
To compare with the analytic results  100,000 flat phase space events has been generated 
for the $2\rightarrow (n-2)$ gluon scattering. 
The events are required to have  cuts in order to avoid soft and collinear
regions in the  momenta of  the outgoing gluons.  
The evaluation time for 10,000 events is: for a $2\rightarrow 2$ gluon ordered helicity amplitude 9 seconds,
for a $2\rightarrow 3$ gluon ordered helicity amplitude 35 seconds and for a
for a $2\rightarrow 4$ gluon ordered helicity amplitude 107 seconds.
Note that using the integration-by-parts method of ref.~\cite{Ellis:2006ss} 
the evaluation time for 10,000 events would be approximately 90,000 second.
The six-gluon evaluation is only three times slower than the five gluon
evaluation and eleven times slower than the four gluon amplitude.

\section{Summary and Outlook}
The numerical unitarity method provides an efficient method
to evaluate  next-to-leading order  corrections to multi-leg
hard scattering amplitudes. It is applicable for processes
including massive and massless particles as well as bosons and fermions.
Very recently it has been generalized
to $D$-dimensions so  one can reconstruct the full amplitude including the rational part~\cite{GKM}.
We expect that in the future it will be used in a number of important
physics applications.

\appendix

\end{document}